**Machine Learning Guided Discovery of Gigantic Magnetocaloric Effect in $HoB_2$ Near Hydrogen Liquefaction Temperature**

*P.B Castro[A,B], K. Terashima[A], T. D Yamamoto[A], Z. Hou[C], S. Iwasaki[A], R. Matsumoto[A,B], S. Adachi[A], Y. Saito[A,B], P. Song[A,B], H. Takeya[A] and Y. Takano[A,B]*

[A]National Institute for Materials Science, 1-2-1 Sengen, Tsukuba, Ibaraki 305-0047, Japan

[B]University of Tsukuba, 1-1-1 Tennodai, Tsukuba, Ibaraki 305-8577, Japan

[C]State Key Laboratory of Structural Chemistry, Fujian Institute of Research on the Structure of Matter, Chinese Academy of Sciences, Fuzhou 350002, China

Corresponding author:
Pedro Baptista de Castro   E-mail: CASTRO.Pedro@nims.go.jp
Kensei Terashima   E-mail: TERASHIMA.Kensei@nims.go.jp
Yoshihiko Takano   E-mail: TAKANO.Yoshihiko@nims.go.jp

***Abstract***

Magnetic refrigeration exploits the magnetocaloric effect which is the entropy change upon application and removal of magnetic fields in materials, providing an alternate path for refrigeration other than the conventional gas cycles. While intensive research has uncovered a vast number of magnetic materials which exhibits large magnetocaloric effect, these properties for a large number of compounds still remain unknown. To explore new functional materials in this unknown space, machine learning is used as a guide for selecting materials which could exhibit large magnetocaloric effect. By this approach, $HoB_2$ is singled out, synthesized and its magnetocaloric properties are evaluated, leading to the experimental discovery of gigantic magnetic entropy change 40.1 J $kg^{-1}$ $K^{-1}$ (0.35 J $cm^{-3}$ $K^{-1}$) for a field change of 5 T in the vicinity of a ferromagnetic second-order phase transition with a Curie temperature of 15 K. This is the highest value reported so far, to our knowledge, near the hydrogen liquefaction temperature thus it is a highly suitable material for hydrogen liquefaction and low temperature magnetic cooling applications.

***Introduction***

The magnetocaloric effect (MCE) is becoming a promising approach into environmental friendly cooling, as it does not depend on the use of hazardous or greenhouse gases[1–4] while being in principle able to attain a higher thermodynamic cycle efficiency[1,5,6] where this cycle makes use of the magnetic entropy change ($\Delta S_M$) and the adiabatic temperature change ($\Delta T_{ad}$) through the application/removal of a magnetic field in a material. Since large values of $\Delta S_M$ are usually achieved near the magnetic transition temperature ($T_{mag}$), the working temperature range is confined around the $T_{mag}$ of the material. MCE was first used to achieve ultra-low cryogenic temperatures (below 1 K)[7] and has been widely used for liquefying He[8] where the main component is the gadolinium gallium garnet $Gd_3Ga_5O_{12}$(GGG)[9]. The remarkable discovery of giant MCE near room temperature in materials such as $Gd_5Si_2Ge_2$[10], $La(Fe,Si)_{13}$[11] and $MnFeP_{1-x}As_x$[3] families, shifted the main focus of research into finding and tuning new materials, such as NiMnIn Heusler alloys[12], working around this temperature range due to its potential economic and environmental impact.[2]

On the other hand, there is an increasing demand for cooling systems around hydrogen liquefaction temperature ($T$ = 20.3 K), since liquid hydrogen is one of the candidates as green fuel for substitution of petroleum-based fuels[13] and also widely needed as a rocket propellant and space exploration fuel[14,15]. It has been shown that MCE based refrigerators prototypes have shown to be highly appropriate for this task.[5] In this context, the discovery of materials exhibiting remarkable MCE response near the liquefaction temperature of hydrogen is highly anticipated.

One way of tackling such a problem is by taking advantage of data-driven approaches, such as machine learning (ML), as it has been successfully applied from the modeling of new superconductors[16,17] and thermoelectric[18,19] to the prediction of synthesizability of inorganic materials[20]. However, in the case of MCE, this kind of approach has not been extensively tried, being limited to first principle calculations which have been restricted to non rare-earth systems.[21]

As a result of extensive research in magnetocaloric materials, accumulated data about MCE properties of diverse types of materials can be accessed in recent reviews[2,22]. In addition, recent efforts into extracting the $T_{mag}$ of materials from research reports led to the creation of the MagneticMaterials[23], an auto-generated database of magnetic materials built by natural language processing which contains a vast number of materials where their magnetic properties are known. Among these materials, there are still many in which its MCE properties have not been evaluated. Therefore, by combining the known and unknown data, a data-driven trial can be used as a guide for finding materials with high MCE response.

In this work, we attempt a novel approach by using ML to screen and select compounds which might exhibit a high MCE performance, focusing on ferromagnetic materials with Curie temperature ($T_C$) around 20 K. For this purpose, we collected data from the literature[2,23] in order to train an ML algorithm for an attempt to predict $\Delta S_M$ of given material composition. By this method, we singled out $HoB_2$ ($T_C$ = 15 K[24]) as a possible candidate, leading us to the experimental discovery of a giant MCE of $|\Delta S_M|$ = 40.1 J kg$^{-1}$ K$^{-1}$ (0.35 J cm$^{-3}$ K$^{-1}$) for a field change of 5 T in this material.

***Materials and Methods***

***Data Acquisition and Machine Learning Model Building***

Figure 1 shows the schematic flow exhibiting the construction of the machine-learned model for MCE materials. We started with the screening of magnetocaloric relevant papers from the MagneticMaterials[23] database and gathered the reported MCE properties contained therein, mostly focusing on the reported values of $|\Delta S_M|$ for a given field change ($\mu_0\Delta H$) of given material composition. Combining the obtained data with the data reported by a recent review,[2] 1644 data points were obtained. For an attempt at the prediction of the MCE property of a material, namely $|\Delta S_M|$, it was chosen to use compositional based features (such as the atomic mass of constituent elements, their specific heat at 295 K, number of valance electrons, etc.), which was directly extracted from the material compositions by using the XenonPy package[25]. Combining the extracted features with the reported values of $|\Delta S_M|$ and $\mu_0\Delta H$ a gradient boosted tree algorithm implemented on the XGBoost[26] package was trained over 80% of the total data. To further improve the prediction power, model selection, and hyperparameter tuning was done by using a Bayesian optimization technique implemented into the HyperOpt[27] package by minimizing the mean absolute error (MAE) 10 fold cross-validation score. The resulting model achieved an MAE of 1.8 J kg$^{-1}$ K$^{-1}$ while tested on the remaining 20% of the data. For more details about the data acquisition, data processing and model building see the Supplementary Information sections 1-3.

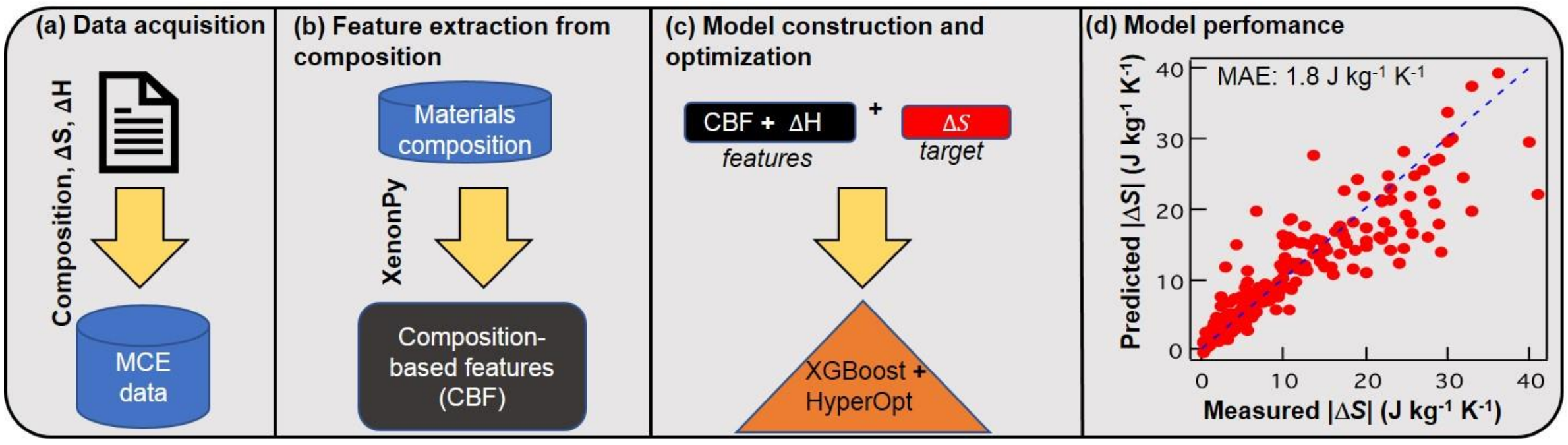

**Figure 1.** Diagram for the construction of the machine learning model for MCE. (a) As a first step, magnetocaloric relevant papers were screened from the MagneticMaterials[23] database followed by the extraction of MCE data from these papers and also from a previous review[2]. (b) Features based on compositions were extracted by the XenonPy[25] python package and used in conjunction with the reported values of field change. (c) Model optimization was achieved using the HyperOpt package by minimizing the MAE score. (d) Model performance was evaluated by the comparison of the predicted $\Delta S_M$ values obtained from the constructed model with those of reported ones for approximately 300 unseen materials to the model, where an MAE of 1.8 J kg$^{-1}$ K$^{-1}$ was achieved.

After the model construction, we examined 818 unknown $\Delta S_M$ text-mined compositions with $T_C \leq 150$ K contained in the MagneticMaterials using the following criteria: the predicted value of $|\Delta S_M|$ higher than 15 J kg$^{-1}$ K$^{-1}$, alloys only, chemical composition containing heavier rare earth elements (Gd-Er) and free of toxic elements, such as arsenic. As a result, $HoB_2$ ($AlB_2$ type, space group $P6/mmm$) was selected, for having the highest predicted $|\Delta S_M|$ (16.3 J kg$^{-1}$ K$^{-1}$) for $\mu_0\Delta H$ = 5 T among the binary candidates followed by its synthesis and characterization of its MCE properties.

## Sample Synthesis

Polycrystalline samples of $HoB_2$ were synthesized by an arc-melting process in a water-cooled copper heath arc furnace. Stoichiometric amounts of Ho (99.9% purity) and B (99.5% purity) were arc melted under Ar atmosphere. To ensure homogeneity, the sample was flipped and melted several times followed by annealing in an evacuated quartz tube at 1000 $^{o}$ C for 24 hours and it was finally water quenched. X-ray diffraction was carried out and $HoB_2$ was confirmed as the main phase structure of the obtained sample (Fig. S2a).

## Magnetic Measurements

Magnetic measurements were carried out by a superconducting quantum interference device (SQUID) magnetometer contained in the MPMS XL (Magnetic Property Measurement System, Quantum Design)

## Specific Heat Measurement

Specific heat measurement was carried out in a PPMS (Physical Property Measurement System, Quantum Design), equipped with a heat capacity option.

## Results

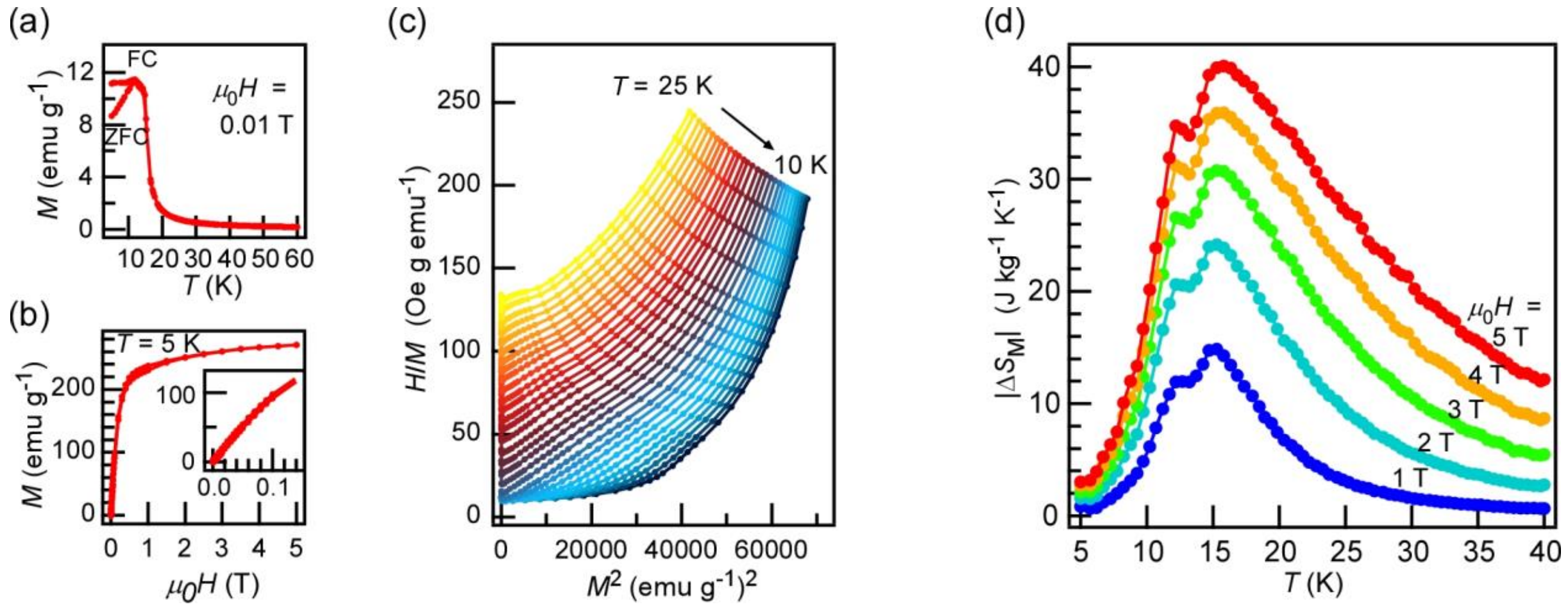

**Figure 2**. Magnetization measurements for $HoB_2$. (a) zero-field cooling (ZFC) and field cooling (FC) *M-T* curves for an applied field of $\mu_0 H = 0.01$ T. (b) *M-H* curve at $T = 5$ K. Inset: Lower field range. (c) Arrott plots deduced from M-T curves (Figure S1a), showing a positive slope for all curves indicating a SOPT. (d) Magnetic entropy change calculated from *M-T* curves shown in Figure S1a using Equation (1). For a field change of 5 T, $|\Delta S_M|$ peaks at 40.1 J $kg^{-1}$ $K^{-1}$ around $T$~15 K.

Figure 2a shows the isofield magnetization (*M-T*) curve of the synthesized polycrystalline $HoB_2$ for an applied field of 0.01 T. $HoB_2$ orders ferromagnetically at $T_C = 15$ K without thermal hysteresis, consistent with previous reports[24]. The isothermal magnetization (*M-H*) at $T$ = 5 K, shown at Fig. 2b, reveals a negligible magnetic hysteresis at $HoB_2$, a common characteristic of second-order phase transition materials (SOPT)[1,28]. This fact is also further confirmed by the Arrott plots shown in Fig. 2c wherein all slopes are positive, in accordance with the Banerjee criterion[29] for SOPT materials, and by building the universal scaling curve as proposed by earlier works[30] (Fig. S4) . A vast number of $M$-$T$ curves for fields ranging from 0-5 T were measured (Fig. S1a) in order to evaluate $|\Delta S_M|$ (Fig. 2d) using the Maxwell relation:

$$\Delta S_M = \mu_0 \int_0^H \left(\frac{\partial M}{\partial T}\right)_H dH \; (1)$$

For a field change of 5 T, we obtained a$|\Delta S_M|$ = 40.1 J kg$^{-1}$ K$^{-1}$ in the vicinity of $T_C$.

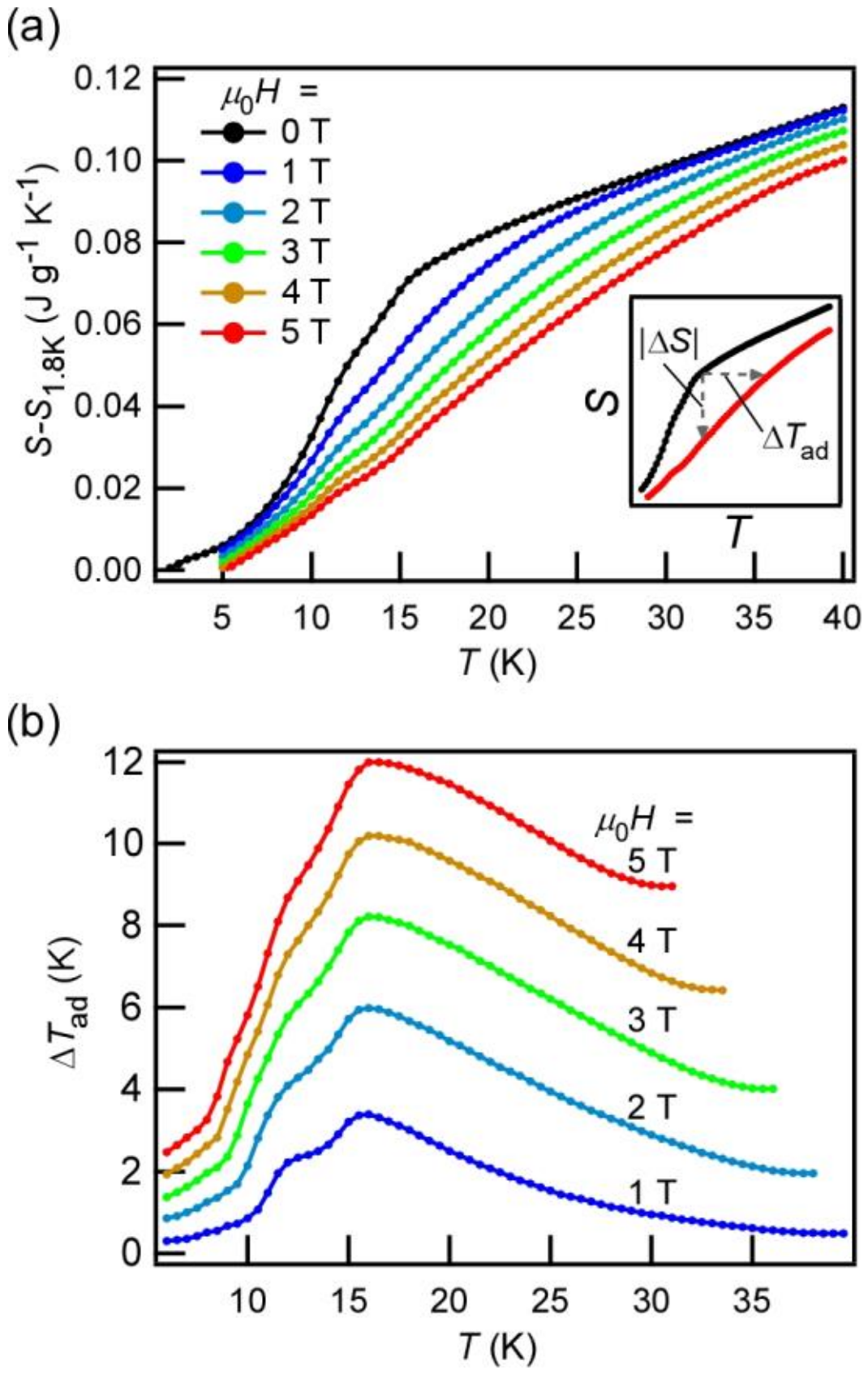

**Figure 3.** Calorimetric properties of $HoB_2$. (a) Entropy as a function of temperature obtained from the zero-field specific heat measurement (Figure S1b). Entropy curves under magnetic fields are obtained combining the entropy curve at 0 T with $|\Delta S_M|$ values of Figure 2(d). Inset shows the definition of $\Delta S$ and $\Delta T_{ad}$. (b) $\Delta T_{ad}$ calculated from (a).

For further evaluation of the MCE performance of $HoB_2$, specific heat measurement was carried out (Fig. S1b) to calculate the entropy curves (see Supplementary Information for the details) and the adiabatic temperature change ($\Delta T_{ad}$) was obtained (Fig. 3b**)**, revealing a maximum $\Delta T_{ad}$ of 12 K for the field change of 5 T.

**Discussion**

In order to compare the performance of $HoB_2$ with other candidates for refrigeration applications near hydrogen liquefaction temperature, such as $ErAl_2$[5], representative large $\Delta S_M$ (for $\mu_0 \Delta H$ = 5 T) materials around $T$ = 20 K are displayed in Table 1. We also show the values of $|\Delta S_M|$ in units of J $cm^{-3}$ $K^{-1}$, which is the ideal unit for the application point of view.[6,22] For a more comprehensive comparison of materials between liquid helium and liquid nitrogen range, see Supplementary Information section 8.

Except for single-crystalline $ErCo_2$ which exhibits a first-order phase transition (FOPT), $HoB_2$ manifests the largest $|\Delta S_M|$ (in both J $kg^{-1}$ $K^{-1}$ and J $cm^{-3}$ $K^{-1}$) and $\Delta T_{ad}$ for a field change of 5 T around hydrogen liquefaction temperature. Among SOPT materials, it also exhibits the largest volumetric entropy change ($\Delta S_M$ in J $cm^{-3}$ $K^{-1}$) in the temperature range from liquid helium (4.2 K) to liquid nitrogen (77 K) (see Fig. S5). It is important to emphasize that, this gigantic magnetocaloric effect is observed in the vicinity of a SOPT. SOPT materials have the advantage of being free of magnetic and thermal hysteresis while having broader $\Delta S_M$ peaks. Thus they are likely to be more suitable for refrigeration purposes than FOPT which tend to be plagued by these problems[1,28]. In other words, $HoB_2$ is a high-performance candidate material for low-temperature magnetic refrigeration applications such as hydrogen liquefaction.

| Material | $T_{mag}$ (K) | $\lvert\Delta S_M\rvert$(J kg$^{-1}$ K$^{-1}$) | $\lvert\Delta S_M\rvert$(J cm$^{-3}$ K$^{-1}$) | $\Delta T_{ad}$ (K) | Transition Type | Reference |
|---|---|---|---|---|---|---|
| **$HoB_2$** | **15** | **40.1** | **0.35** | **12** | **SOPT** | **This work** |
| EuS | 18.2 | 37 | 0.21 | 10.4 | SOPT | 31 |
| $ErAl_2$ | 14 | 36 | 0.22 | 11.1 | SOPT | 32 |
| $ErCo_2$ | 30 | 36 | 0.37 | 9.5 | FOPT | 33 |
| TmGa | 15 | 34.2 | 0.30 | 9.1 | SOPT | 34 |
| $HoAl_2$ | 27 | 28.8 | 0.17 | * | SOPT | 35 |
| $GdCoC_2$ | 15 | 28.4 | 0.23 | * | SOPT | 36 |
| HoN | 18 | 28.2 | 0.29 | * | SOPT | 37 |
| $HoNi_2$ | 13.9 | 26.1 | 0.27 | 8.7 | SOPT | 38 |
| ErFeSi | 22 | 23.5 | 0.18 | 7.1 | SOPT | 39 |

**Table 1** Comparison of MCE-related properties in $HoB_2$ and other materials exhibiting large magnetocaloric response around the liquefaction temperature of hydrogen for field change of 5 T. The data was taken from the references[31–39] in J kg$^{-1}$ K$^{-1}$ and also converted into J cm$^{-3}$ K$^{-1}$ by using the ideal density according to the AtomWork[40] database. Asterisk (*) indicates an unreported value.

## Conclusions

In summary, by using a machine-learning aided approach, we have successfully unveiled a ferromagnet that would manifest a high magnetocaloric performance with a transition temperature around the hydrogen liquefaction temperature. By synthesizing and evaluating its MCE properties, we discovered a gigantic magnetocaloric effect of $HoB_2$ in the vicinity of a SOPT at $T_C$ = 15 K, where the maximum obtained magnetocaloric entropy change was 40.1 J kg$^{-1}$ K$^{-1}$ (0.35 J cm$^{-3}$ K$^{-1}$) with an adiabatic temperature change of 12 K for a field change of 5 T, the highest value reported until now, to our knowledge, for materials working near liquefaction temperature of hydrogen.

**Acknowledgments**
This work was supported by the JST-Mirai Program Grant No. JPMJMI18A3, JSPS KAKENHI Grant No. 19H02177, JST CREST Grant No. JPMJCR16Q6. P. B. Castro also acknowledges the scholarship support from the Ministry of Education, Culture, Sports, Science and Technology (MEXT), Japan.

**Authors contributions**
Y T conceived the project idea. P B C did the data acquisition, preprocessing, machine learning model building and material predicting with assistance of H Z. P B C and K T did the sample synthesis with assistance of H T. K T did the magnetization measurements and heat capacity measurements with T D Y. K T and P B C analyzed the experimental data. P B C and KT wrote the manuscript. All authors discussed the manuscript together.

**Conflicts of Interest**

The authors declare no conflict of interest

**Supplementary Information** is available for this paper at NPG Asia Materials website

# Supplementary Information

## Machine Learning Guided Discovery of Gigantic Magnetocaloric Effect in $HoB_2$ Near Hydrogen Liquefaction Temperature

*P.B Castro[A,B], K. Terashima[A], T. D Yamamoto[A], Z. Hou[C], S. Iwasaki[A], R. Matsumoto[A,B], S. Adachi[A], Y. Saito[A,B], P. Song[A,B], H. Takeya[A] and Y. Takano[A,B]*

[A]National Institute for Materials Science, 1-2-1 Sengen, Tsukuba, Ibaraki 305-0047, Japan

[B]University of Tsukuba, 1-1-1 Tennodai, Tsukuba, Ibaraki 305-8577, Japan

[C]State Key Laboratory of Structural Chemistry, Fujian Institute of Research on the Structure of Matter, Chinese Academy of Sciences, Fuzhou 350002, China

Corresponding author:
Pedro Baptista de Castro   E-mail: CASTRO.Pedro@nims.go.jp
Kensei Terashima   E-mail: TERASHIMA.Kensei@nims.go.jp
Yoshihiko Takano   E-mail: TAKANO.Yoshihiko@nims.go.jp

This supplementary information consists of 8 sections. The first three sections (1-3) will be focused on describing the details of the data acquisition and model building procedure. Here we will not show any mathematical description of the machine learning model, but we will cite the corresponding references.

The following four sections (4-7) will be focused on discussing experimental data. In the fourth section, we show the *M-T* curves and the specific heat data used for obtaining $|\Delta S_M|$ and the entropy curves of the main text sample (hereafter labeled as Sample **#1**). In the fifth, we make a brief discussion regarding the reproducibility of $|\Delta S_M|$ value as well as XRD pattern from sample to sample. For this purpose, a second sample was synthesized (hereafter labeled as Sample **#2**). The data of Sample **#2** is also used to check the validity of the entropy curve deduced from the magnetization measurements in the main text, and the result is shown in the sixth section where we compare the magnetocaloric effect calculated from both magnetization and specific heat data. At the seventh, we discuss the nature of the magnetic transition by examining the universal scaling as suggested by Refs[1–4] using the $|\Delta S_M|$ data of Sample **#1**. In the last section (8), we will compare the obtained value of $\Delta S_M$ (J cm$^{-3}$ K$^{-1}$) with representative materials with $T_{mag}$ in the range from liquid helium (4.2 K) to liquid nitrogen (77 K) and whose $\Delta S_M$ is greater than 0.15 J cm$^{-3}$ K$^{-1}$.

# Section 1 - Data Acquisition and Processing

In order to obtain the data needed for training the machine learning model, two different sources were used: the text-mined autogenerated database of magnetic materials, MagneticMaterials[5], and the recent report[6] by Franco *et. al* which contained a comprehensive table of magnetocaloric materials, wherein the values of $\Delta S_M$ and $T_{mag}$ were promptly available.

For the case of the MagneticMaterials database, we screened out magnetocaloric papers by filtering the transition temperature with values lower than 100 K and by filtering the journal titles using the following keywords: magnetocaloric, magnetic refrigeration, cryogenic, caloric, refrigeration resulting in 219 total different journal titles from where the magnetocaloric properties were manually extracted.

To remove any possible duplicates in the final dataset, the materials which contained more than one value of $\Delta S_M$ for a given $\Delta H$ had their values averaged and this average was used as the final value. Lastly, we selected the data within the range of $\Delta H \leq 5$ T, for compatibility with our experimental setup. This procedure gave us the final 1644 data points used for the model construction

# Section 2 - Feature Construction

To obtain the physical properties for the modeling of $\Delta S_M$, the XenonPy[7] package was used. For this, first the materials composition were transformed into the appropriate format taken by XenonPy with the aid of the pymatgen[8] Python package. Namely, the material composition was converted into a python dictionary where the keys are the atomic compositions and the values are their amount. For example, the suitable input in the case of $HoB_2$ would be the python dictionary: {"Ho": 1.0, "B": 2.0}. Given this input, the compositional based features can be then automatically calculated by XenonPy from 58 element-level properties. Below, we list a few examples of the element-level properties:

- **lattice constant**: bulk atomic material lattice constant (for ex. bulk Ho lattice constant)
- **specific heat at 295 K**: atomic specific heat (ex. bulk Ho specific heat at 295 K)
- **number of valence p electrons**
- **atomic number**

For the generation of the compositional features for a candidate compound, the element-level properties were combined using 7 different featurizers (included in the XenonPy package) such as:

- **Weighted average**: $f_{ave,i} = w_A^* f_{A,i} + w_B^* f_{B,i}$
- **Weighted variance**: $f_{\sum,i} = w_A f_{A,i} + w_B f_{B,i}$
- **Max-pooling**: $f_{max,i} = max\,(f_{A,i}, f_{B,i})$

Where $w_A^*$ and $w_B^*$ are the normalized composition ($w_A^* + w_B^* = 1$) and $f_{A,i}$ and $f_{B,i}$ are the ith atomic feature of compound A and B. In the case of $HoB_2$, the weighted average of atomic number compositional feature would be calculated as:*

$$atomicnumber_{ave} = \frac{1}{3}(AtomicnumberofHo) + \frac{2}{3}(AtomicnumberofB)$$
$$= \frac{1}{3}(67) + \frac{2}{3(5)} = 25.6667$$

The full list of available element-level properties and featurizers can be found in the package website at: https://xenonpy.readthedocs.io/en/latest/features.html#data-access.

Another feature calculator used was the counting of the 94 atomic elements (from H to Pu, implemented in XenonPy as Counting featurizer). For example, in the case of $HoB_2$, the final features would be given as Ho: 1.0, B:2.0, while all other elements would be zero.

Features which all values were zero, which contained infinite values or not a number (NaN) were removed. In the end, 408 features were used, where 407 were generated by XenonPy and the last was the field change (ΔH). This workflow is summarized below:

| total used features | | atomic features | | compositional math calculation | | atomic elements $^{1}H$ - $^{94}Pu$ | | NaN infinite zero | | ΔH |
|---|---|---|---|---|---|---|---|---|---|---|
| 408 | = | 58 | × | 7 | + | 94 | − | 93 | + | 1 |

# Section 3 - Model Selection and Building

As a first attempt for building the machine learning model, two different models were first tried: a LASSO[9] (least absolute shrinkage and selection operator) linear model implemented on the scikit-learn[10] python package and the gradient boosted tree implemented into the XGBoost[11] package. In order to compare their performance the three following metrics were used: coefficient of determination ($R^2$), root mean squared error (RMSE) and mean absolute error (MAE), which are defined as:

$$R^2 = 1 - \sum_{i}^{n} (y_i - f_i)^2 \Big/ \sum_{i}^{n} \left(y_i - y_{avg}\right)^2$$

$$MAE = \frac{1}{n}\sum_{i}^{n} |f_i - y_i|$$

$$RMSE = \sqrt{MSE}; MSE = \frac{1}{n}\sum_{i}^{n} (f_i - y_i)^2$$

where $y_i$ is the measured value and $f_i$ is the predicted value. Each model metrics is shown in Table 1. The training and testing of these models were done using the standard machine learning practices: the training/testing data set were split into an 80%/20% ratio using the scikit-learn package, the model was trained in the training data and the results were done in the testing data.

| | LASSO | XGBoost |
|---|---|---|
| $R^2$ | 0.33 | 0.85 |
| MAE | 4.7 | 1.78 |
| RMSE | 6.8 | 3.21 |

**Table S1:** Metrics values for two different machine learning models for the testing set.

As we can see from Table **S1**, LASSO does not perform well, indicating that this problem is a highly non-linear one. In the present work, we decided to use XGBoost as it has been shown to have great performance in diverse machine learning problems, being robust and widely used in production. The complete mathematical description of these algorithms can be found at Ref.[9] for LASSO and Ref. [11] for XGBoost.

In the XGBoost model, there are several important hyperparameters that control how the tree model is constructed and trained. The amount of trees (denoted as n_estimators) subsample for bagging (subsample ratio of the training instance, denoted as subsample), subsample ratio of features(denoted as colsample_bytree), maximum depth of a tree (denoted as max_depth), minimum number of instances needed in each node (denoted as min_child_weight) and so on. The complete list of parameters and their definitions can be found at the XGBoost documentation at: https://xgboost.readthedocs.io/en/latest/parameter.html.

For the selection of suitable hyperparameters, we tried exhaustive random and grid search, implemented at the scikit-learn package, and the python package HyperOpt[12]. HyperOpt implements a bayesian approach optimization which usually can find local/global minimums faster than the exhaustive approaches. We found that in our case, the hyperparameters found by using the HyperOpt package for optimizing the MAE in a 10 fold cross-validation procedure of the training set gave the best results compared to random and grid search. Cross-validation is one of the standard techniques used to test the effectiveness of a machine learning model and its in-depth explanation can be found in the scikit-learn documentation and in freely available machine learning books[9].

The hyperparameters found by HyperOpt used for the final model are shown in Table S2**.** By using these parameters, the final $R^2$ score was 0.85 while MAE was 1.78 for the testing set, and this model was used for the predictions of $\Delta S_{M}$;

| N_estimators | Subsample | Colsample_bytree | Max_depth | Min_child_weight | Learning_rate |
|---|---|---|---|---|---|
| 720 | 0.75 | 0.5 | 8 | 5.0 | 0.05 |

**Table S2:** Final tuned hyperparameters found by the HyperOpt package.

# Section 4 - Magnetization and specific heat of Sample #1

Figure S1a shows the magnetization data for various applied fields in the range from 0 to 5 T for Sample **#1**. The magnetic entropy change ($\Delta S_M$) shown in Figure 2d of the main text was calculated from this data through the numerical integration of $\partial M/\partial T(T, H)$. Figure S1b shows the specific heat data of Sample #1 under zero field. The entropy curve under zero-field shown in Figure 3a of the main text was calculated by using this specific heat data. Note that in addition to a peak at $T_C$ ~15 K, a second peak appears around 11 K in the specific heat data. This is probably due to a spin reorientation transition as discussed in a related compound $DyB_2$[13]. At this stage, we keep its physical origin as an open question to be clarified in future work, since it is not the main focus of present work.

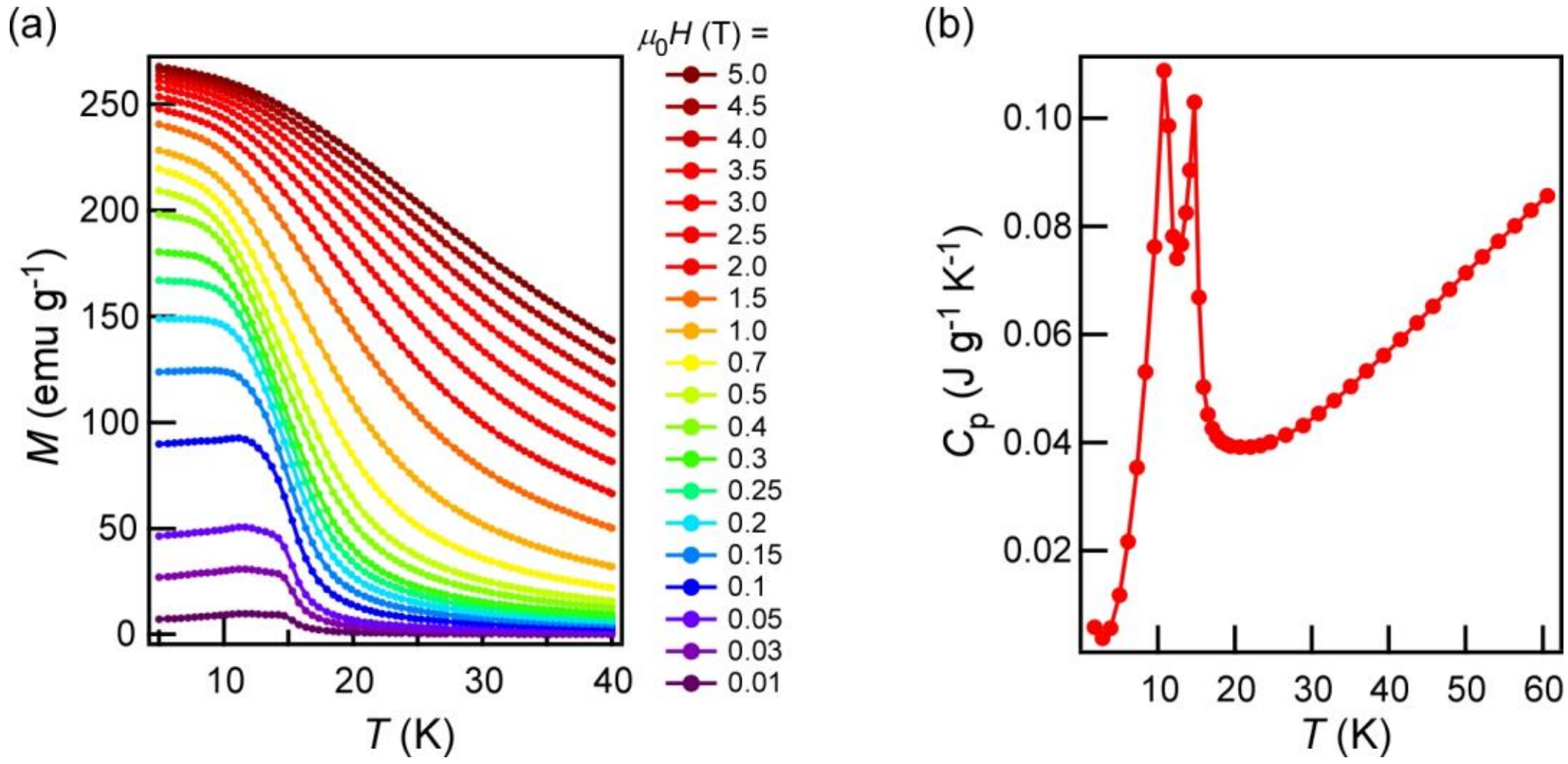

**Figure S1.** (a) *M-T* curves of $HoB_2$ for applied fields ranging from 0 to 5 T. (b) Specific heat data of $HoB_2$ from 1.8 to 60 K.

# Section 5 - Reproducibility

In order to evaluate the reproducibility of our findings, Sample **#2** was synthesized by using the same arc melting procedure as in the main text. The XRD patterns of both samples are shown in Figure S2a. For both samples, $HoB_2$ is the majority phase (~95 %) while there is a slight difference in the observed impurity peaks. This difference might be associated with the presence of a small minority phase of unreacted Ho in Sample **#1** while Sample **#2** exhibits an impurity peak associated with $HoB_4$. For comparison, we have also evaluated $|\Delta S_M|$ in Sample **#2** as shown in Figure S2b. While Sample **#1** peaks at $|\Delta S_M|(\Delta H = 5$ T, $T = 15$ K$)| = 40.1$ J kg$^{-1}$ K$^{-1}$, Sample **#2** peaks at 39.1 J kg$^{-1}$ K$^{-1}$, showing the reproducibility of the gigantic magnetocaloric effect in this material. The small difference in $|\Delta S_M|$ value between Sample **#1** and **#2** is probably due to the difference in the impurity phase contained in each sample.

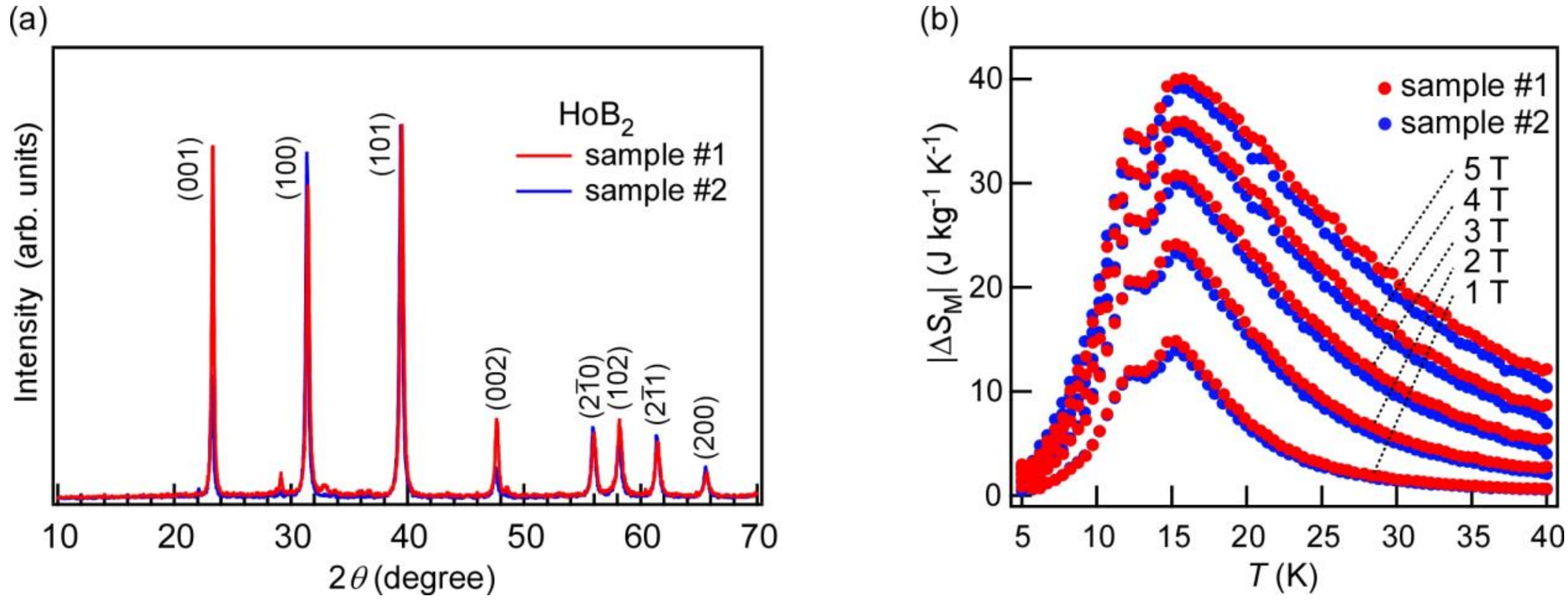

**Figure S2**. (a) XRD pattern for both Sample **#1** and **#2**. (b) Reproducibility of magnetocaloric effect in $HoB_2$ for two different samples

# Section 6 – Entropy from specific heat measurements and simulated curve under field from magnetization measurements

In this section, we show the comparison of the entropy curve under magnetic field deduced from two experimental methods. One is the entropy obtained from specific data taken at $\mu_0 H$ = 0, 2, 5 T for Sample **#2** using the following equation:

$$S(T) = \int_{T_{min}}^{T} \frac{C}{T}\, dT$$

where $T_{min}$ is 1.8 K, and the obtained entropy curves are shown as red circles in Figure S3. Another is the simulated entropy under field (blue circles in Figure S3), that is obtained simply adding $|\Delta S_M|$ evaluated by *M-T* curves to the 0 T entropy. As we can see in Figure S3, both show a good agreement with each other, ensuring a gigantic magnetocaloric effect in $HoB_2$. We have found that this simulation to obtain the entropy curve under field is efficient and fast. Since we were working within a limited machine operation time constraint, we used the same simulation approach to obtain the entropy under field shown in Figure 3(a) of the main text.

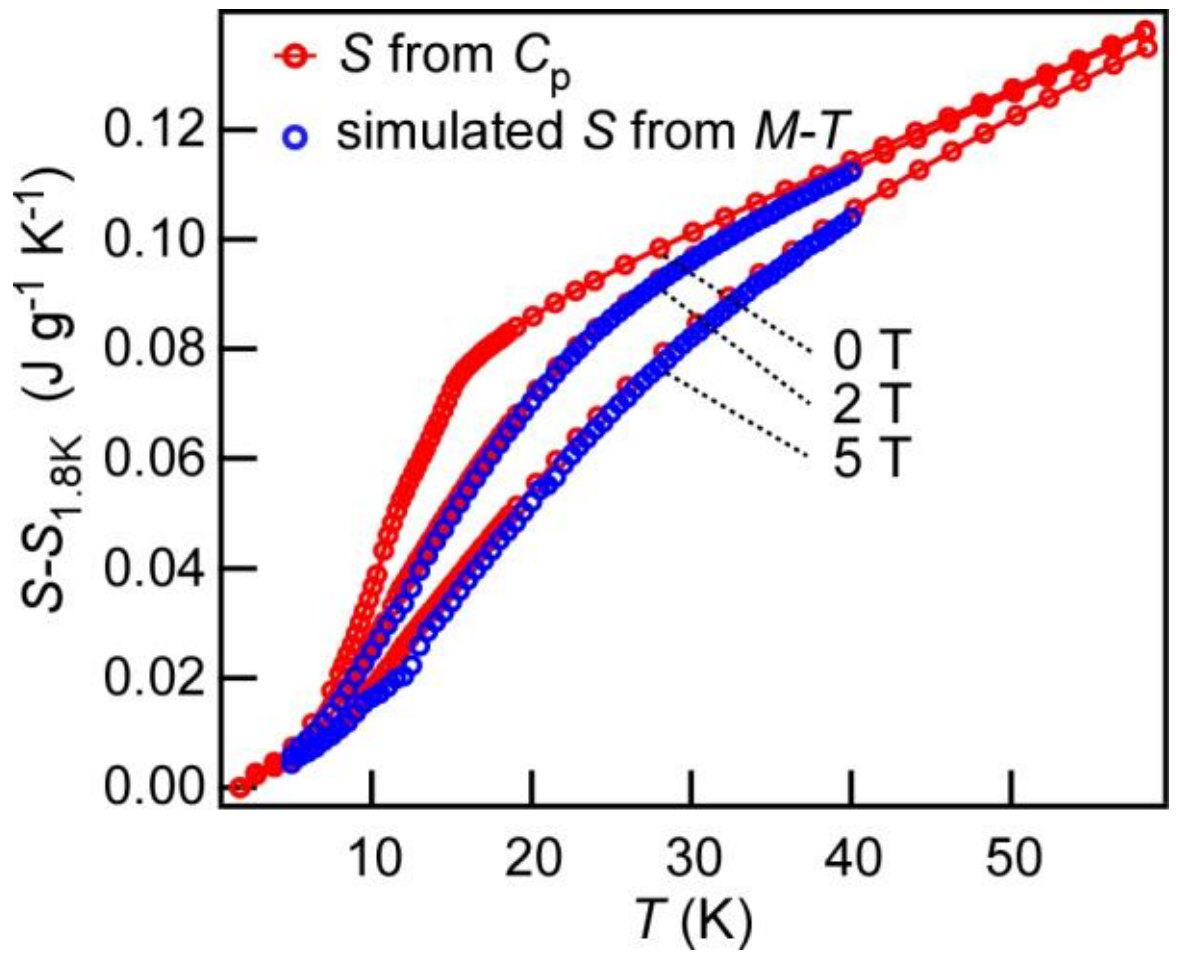

Figure **S3.** Entropy as a function of temperature obtained from the specific heat (red circle) and those by combining the specific heat at 0 T with $|\Delta S_M|$ from magnetization data (blue circle) for Sample **#2**

# Section 7 - Universal scaling of ΔS curve and order of magnetic transition.

To further understand the nature of the magnetic transition at $T_C$ = 15 K, we've examined a so-called universal scaling curve proposed in Refs.[1–4] that depicts the normalized entropy change $|\Delta S_M|/|\Delta S_M^{peak}|$ as a function of normalized temperature $\theta$, defined as:

$$\theta = \begin{cases} -(T - T_C)/(T_{r_1} - T_C), & T \le T_C \\ (T - T_C)/(T_{r_2} - T_C), & T > T_C \end{cases}$$

where $T_{r_1}$ and $T_{r_2}$ are reference temperatures at which $|\Delta S_M|/|\Delta S_M^{peak}|$ = 0.5.

For the case of a second-order transition, these curves are expected to collapse into each other, exhibiting a universal behavior, while for the first-order transition, there would be a dispersion between the curves (See for eg. Ref.[4]). As can be seen in Figure **S4**, all the curves for different fields fall into the same universal curve. The divergence around $\theta = -1$ is due to the presence of a second magnetic transition at a lower temperature, similarly observed for other materials exhibiting more than one transition[14,15]. This universal curve suggests that the nature of the magnetic transitions in $HoB_2$ are indeed of second-order, although further experiments would be required to conclude the nature and origin of the second peak at lower temperature regime.

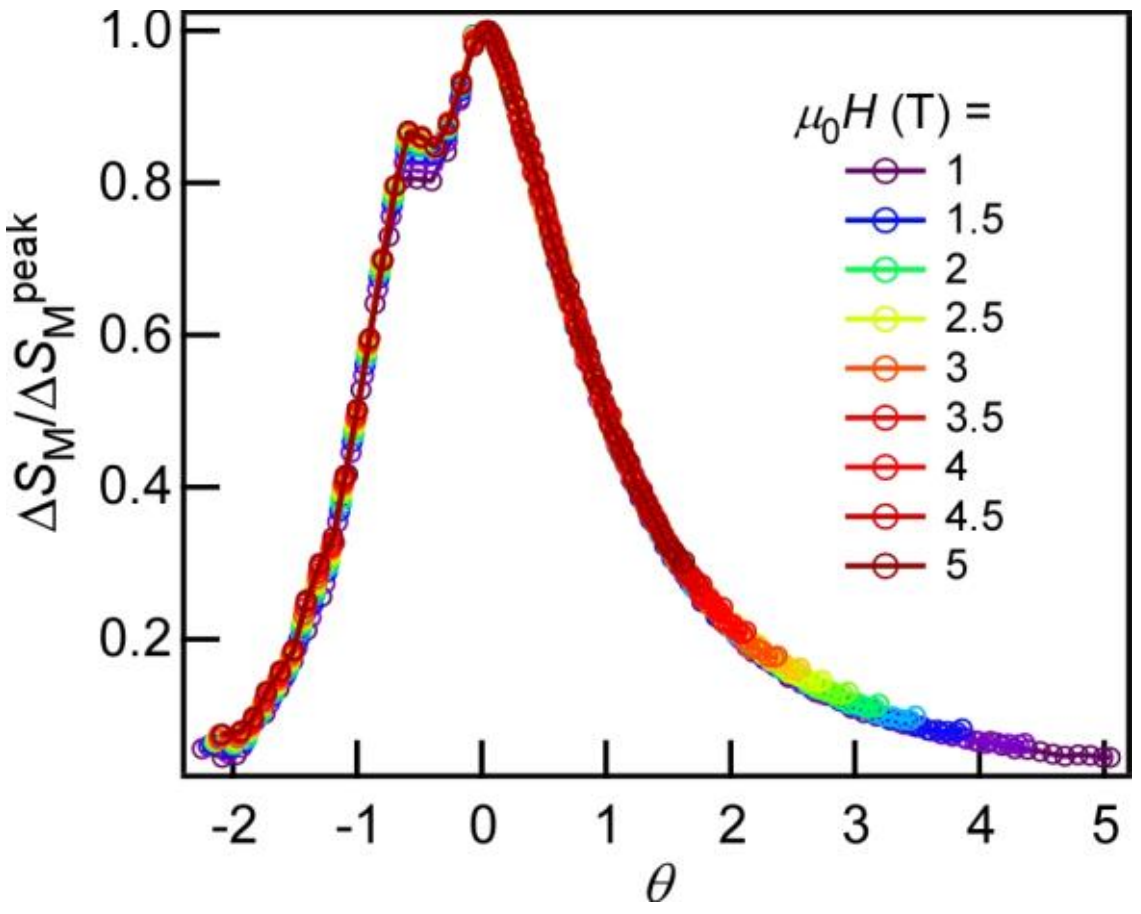

Figure S4. Universal scaling curve of $|\Delta S_M|$ in $HoB_2$, obtained from the data in Figure 2(d) of the main text.

# Section 8 – Comparison of volumetric entropy change at low temperatures range for representative materials.

Figure S5 shows diverse magnetocaloric materials enumerated by $T_{mag}$ in the temperature range of liquid helium (4.2 K) to liquid nitrogen (77 K). These materials were selected from our database for exhibiting a $|\Delta S_M| \geq 0.15$ J cm$^{-3}$ K$^{-1}$ for $\mu_0 \Delta H = 5$ T. In here we emphasize again the importance of comparison in the volumetric unit, as it is the most appropriate unit when considering the material as a candidate for applications such magnetic refrigeration[16,17].

$HoB_2$, exhibits the highest volumetric $|\Delta S_M|$ for all SOPT materials in the temperature range of 4.2 K to 77 K, being comparable to FOPT materials such as $ErCo_2$ (#27 in the figure) and $Gd_5(Si_{0.33}Ge_{3.67})$ (#35,37 in the figure). Furthermore, its closeness to the hydrogen liquefaction temperature (indicated by the dashed line) combined with this gigantic volumetric $\Delta S_M$ solidifies it as a remarkable candidate for magnetic coolant especially aimed at the hydrogen liquefaction stage and low-temperature applications.

**Figure S5**. $|\Delta S_M|$ (J cm$^{-3}$ K$^{-1}$) ($\mu_0 \Delta H$ = 5 T) of representative materials at the liquid helium

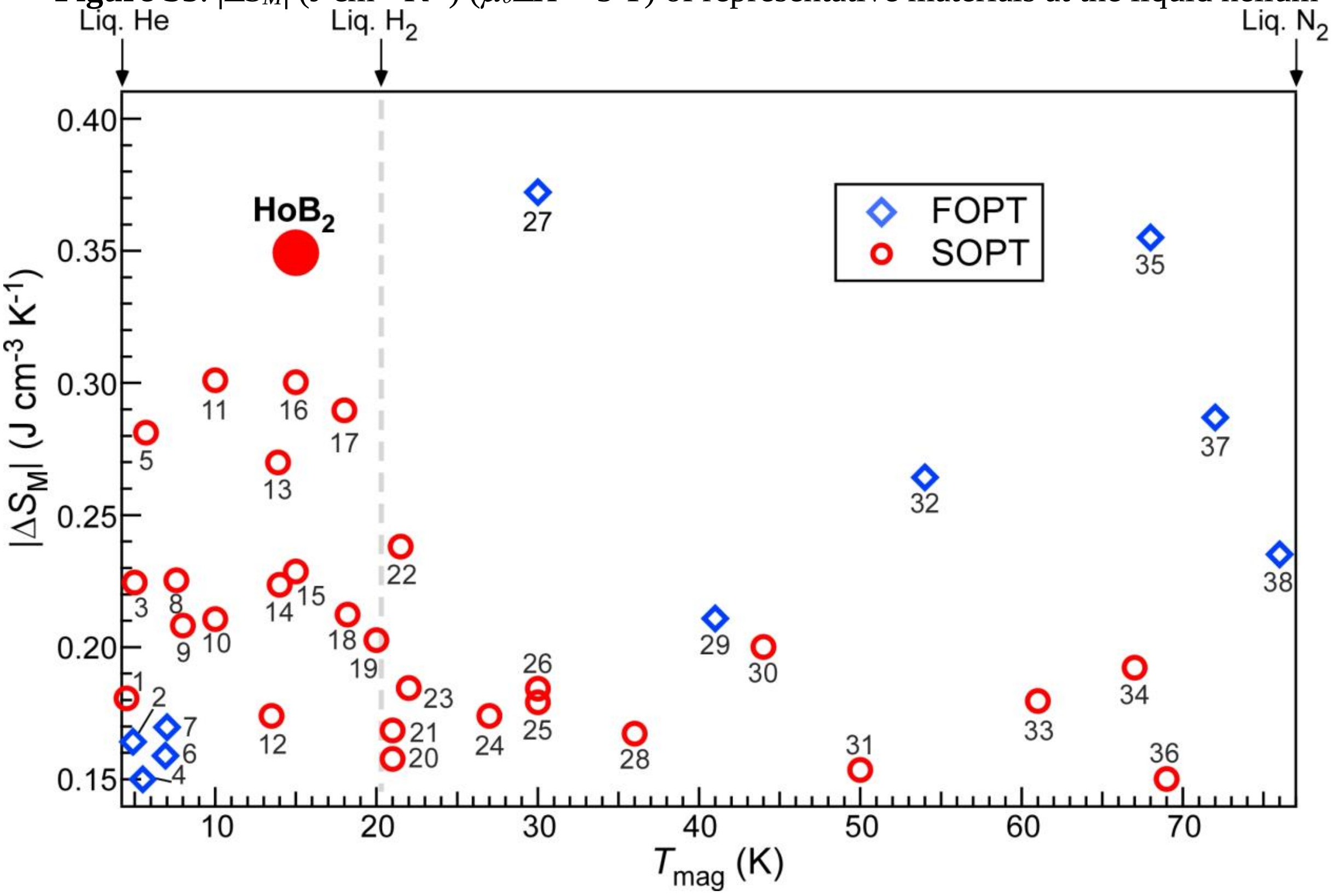

(4.2 K) to liquid nitrogen (77 K) temperature range. The dashed line indicates the hydrogen liquefaction temperature. The blue triangles indicate materials that undergo a FOPT while the red dots a SOPT. The materials compositions are indicated by the numbers in the table below. The conversion to volumetric units used the density available at the AtomWork[18] database unless otherwise specified by the authors.

| | | | | |
|---|---|---|---|---|
| 1: $ErMn_2Si$[19] | 9: $Ho_2PdSi_3$[20] | 17: HoN[21] | 25: $Ho_{12}Co_7$[22] | 32: $Gd_3Ru$[23] |
| 2: $HoNi_2Si_2$[24] | 10: HoPd[25] | 18: EuS[26] | 26: $Ho_2Cu_2Cd$[15] | 33: DyNi[27] |
| 3: ErNiBC[28] | 11: ErNi[29] | 19: HoNiIn[30] | 27: $ErCo_2$[31] | 34: TbNi[32] |
| 4: $ErRu_2Si_2$[33] | 12: $Er_{12}Co_7$[34] | 20: $Ho_2Au_2In$[35] | 28: HoNi[32] | 35: $Gd_5Si_{0.33}Ge_{3.67}$[16,36] |
| 5: $EuTiO_3$[37] | 13: $HoNi_2$[38] | 21: DyN[39] | 29: $Gd_5Ge_4$[36] | 36: GdNi[32] |
| 6: $DyNi_2Si_2$[24] | 14: $ErAl_2$[40] | 22: $DyNi_2$[41] | 30: TbN[21] | 37: $Gd_5Si_{0.33}Ge_{3.67}$[16,42] |
| 7: $Er_2PdSi_3$[43] | 15: $GdCoC_2$[44] | 23: ErFeSi[45] | 31: $DyB_2$[13] | 38: $HoCo_2$[46] |
| 8: HoCoGe[47] | 16 TmGa[48] | 24: $HoAl_2$[49] | | |